\documentstyle[amsmath,aps,prb,multicol,epsfig,floats,axodraw,amssymb,eqsecnum,graphics]{revtex}
\begin{document}
\draft
\twocolumn[

\hsize\textwidth\columnwidth\hsize\csname @twocolumnfalse\endcsname

\title{Negative scaling dimensions and conformal invariance at the Nishimori point in the $\pm J$ random-bond Ising model}
\author{Florian Merz and J. T. Chalker}
\address{Theoretical Physics, University of Oxford, 1 Keble Road, Oxford  OX1 3NP, UK}
\date{\today}
\maketitle

\begin{abstract}
We reexamine the disorder-dominated multicritical point of the two-dimensional
$\pm J$-Ising model, known as the Nishimori point (NP).
At the NP we investigate numerically and analytically the behavior of the disorder correlator, familiar from
the self-dual description of the pure critical point of the two-dimensional Ising model. 
We consider the logarithmic average and the $q$th moments of this correlator in the ensemble average over 
randomness, for continuous $q$ in the range $0<q<2.5$, and
demonstrate their conformal invariance.
At the NP we find, in contrast to the self-dual pure critical point, that the disorder correlators 
exhibit multi-scaling in $q$ which is different from that of spin-spin correlators and that their scaling dimension
becomes negative for $q>1$ and $q<0$. Using properties on the Nishimori line
we show that the first moment ($q=1$) of the disorder correlator is exactly one for all separations. 
The spectrum of scaling dimensions at the NP is not parabolic in $q$.
\pacs{PACS numbers: 05.70.Jk, 75.10.Nr, 75.40.Mg}
\end{abstract}

]
\vskip2pc

\section{Introduction}

Conformal field theory (CFT) has been a key ingredient in understanding critical models and phase-transitions in two 
spatial dimensions\cite{cft}. It has helped
the exact solution of a number of problems and made it possible to construct a classification of critical points. While
so far the success of CFT has been mainly in its application to pure critical systems, one might hope that it 
will be an equally powerful tool for understanding disorder-dominated critical points in two dimensions, such 
as those in disordered magnets or the integer quantum Hall effect:
It is generally expected that the observables of a random system at criticality exhibit conformal invariance when averaged over disorder,
and the averages are believed to be vacuum expectation values of operators in an appropriate conformal field theory. 
Even if this theory is unknown, the bare assumption of the
existence of an underlying conformal field theory leads to strong constraints which may be tested for particular models. 

One model system which has received renewed attention recently is the
$\pm J$ random-bond Ising model ($\pm J$ RBIM), which (amongst others) has a special symmetry line in 
parameter space, known as the Nishimori line\cite{nishiorg}. It has a disorder-dominated 
multicritical point, the Nishimori point (NP), where this line intersects the phase boundary between a 
ferromagnet and a paramagnet. The NP may be a good candidate for the construction 
of a consistent CFT of a random critical point since it is one of the simplest model critical points at the outset. 
The phase transition in the RBIM is also of interest as a version of the quantum Hall plateau transition, 
since there is an exact mapping\cite{chofisher,merzch,gruzberg} from the RBIM to a network model similar that 
used to represent the latter\cite{chalkcodd}. Furthermore, owing to extensive numerical
and analytic effort invested in studying the $\pm J$ RBIM and the NP in particular, accurate estimates for 
various exponents are available, which can be used to test candidate CFTs. 
At the NP the first few integer $q$ moments of the spin-spin  ($\sigma\sigma$-) correlation 
function have been calculated numerically recently\cite{honecker,merzch} and have been demonstrated to obey the 
conformal constraints very accurately. 
The $q$-dependence of their scaling dimension has considerable significance, and one can imagine three possibilities.
If critical behavior at the NP were that of the percolation transition, the dimensions would be independent of $q$.
By contrast, the critical behavior in a pure system gives dimensions increasing linearly in $q$. In fact, at the NP
scaling dimensions of moments of the $\sigma\sigma$- correlator vary non-linearly with $q$.
This behavior is generic in the presence of randomness and referred to as multi-scaling. 
For the RBIM, Read and Ludwig\cite{readlud} considered the expectation values of the disorder 
operator which is the Kramers-Wannier dual to the conventional order operator $\sigma$.
The dual operator will be the main focus of this paper and we will refer to it as 
$\mu$-operator throughout in order to avoid confusion 
between thermal and configurational disorder.
The $\mu$-operators are associated with plaquettes of the Ising lattice: 
the two-point correlator  $\langle\mu(x)\mu(y)\rangle$ 
in a particular distribution of bonds can be represented by choosing an 
arbitrary path connecting the pair of plaquettes at $x$ and $y$ and reversing the 
sign of all the bonds crossed by the path. In this way one arrives at a modified system 
with partition function $Z'$ (compared to the $Z$ of the unmodified model) and the
the correlator is given by the ratio of the two\cite{kadce}, $\langle\mu(x)\mu(y)\rangle=Z'/Z$.
In Ref.\onlinecite{readlud} the authors point out that the $\mu$-operator, is, 
in contrast 
to $\sigma$, not bounded from above anymore when antiferromagnetic exchange interaction are present. 
This opens up the interesting possibility of correlation 
functions which increase with distance and consequently the possibility of negative scaling dimensions, which in turn 
is intimately linked to the central problem of non-unitarity in CFT. The fact that $\mu\mu$-correlators can increase with distance has been
established in Ref.\onlinecite{readlud} for RBIM's with equal concentrations of ferromagnetic and antiferromagnetic exchange interactions.
Here our concern is instead with behaviour at a critical point. The aim of this paper is to explicitly demonstrate 
conformal invariance with negative scaling dimensions by considering the $\mu\mu$-correlator at the NP. One practical 
problem in these calculations is to achieve high sampling while keeping the system size sufficiently large. From this perspective 
making use of the recently developed mapping\cite{merzch} of the RBIM onto the network model for calculating free energies
is crucial since it reduces a formerly exponentially large calculation to one which is only power-law in system size.  

In addition to our numerical results, we show in this paper that the distribution functions of the $\mu$-operators satisfy
strong constraints on the Nishimori line.
In particular we consider the $q$th moments of correlation functions and we show that the moments are
symmetric in $q$ with respect to $q=1/2$. Owing to this symmetry
the $\mu$-operators for $q=1$ average exactly to unity, as discussed in the following section.
Due to the asymmetry between the spin operator and the disorder operator in the presence of bond disorder,
the NP is not microscopically self-dual.
Despite this the question arises whether self-duality is restored in the correlation functions asymptotically for large separations.
The above exact result on the Nishimori line shows that this does not happen, since the spin-spin correlation function decays
with distance.
In Sec.\ref{numerics} we present results from extensive numerical calculations on the correlations of the disorder 
$\mu$-operator for
the $\pm J$ RBIM. We show that the moments of its correlation function obey the conformal constraints thus reinforcing the
idea of an underlying conformal field theory. We find also that the $q$th 
moment of the $\mu\mu$-correlation function increases with separation $r$
for $q>1$, establishing the presence negative scaling dimensions and non-unitarity.
We test our results against previous calculations in the quasi one-dimensional regime and find excellent agreement.
Finally, our results show that multi-scaling at the NP is not parabolic.

\section{Distributions in the RBIM on the Nishimori line}

We consider the two-dimensional nearest-neighbor Ising model on the square lattice with partition
function $Z(\{ J\} ,\beta)=\text{Tr}_{\sigma} \exp [\sum_{\langle ij\rangle} \beta J_{ij}\sigma_{i}\sigma_{j}]$, 
where $\sigma_{i}$ is a classical
spin variable taking the values $\pm 1$ and the $\langle ij\rangle$ denote nearest neighbors.
The exchange couplings $J_{ij}$ are drawn independently from a probability distribution $P(J)$. For general $P(J)$ this
model is known as the random-bond Ising model (RBIM).
In the conventional notation the correlator of a local operator $O_{x}$ is 
\begin{equation}\label{convcorr}
\langle O_{x}O_{y}\rangle=\frac{1}{Z}\text{Tr}_{\sigma}\ O_{x}O_{y} \exp [\sum_{ij} \beta J_{ij}\sigma_{i}\sigma_{j}].
\end{equation}
It was shown by Kadanoff and Ceva\cite{kadce}, however, that if $O$ represents either the order operator or its dual operator
an alternative way of writing Eq.(\ref{convcorr}) is to absorb the product $O_{x}O_{y}$ into
the Hamiltonian by modifying the set $\{ J_{ij} \} \to\{ J'_{ij} \}$ so that
\begin{equation}\label{altref}
\langle O_{x}O_{y}\rangle=Z'/Z,
\end{equation}
where $Z'$ is the partition function evaluated with the set $\{ J'_{ij} \}$.
One important observation when considering averages over bond distributions is that the
properties of the average of $\langle O_{x}O_{y}\rangle$ crucially depend on whether the 
modification $O_{x}O_{y}$ induces takes the set $\{ J_{ij} \}$ out of the random ensemble or not.
To be specific, inserting the order operator $\sigma$ of the Ising model may be viewed as giving the bonds along a 
semi-infinite path on the lattice an imaginary
part while the $\mu$-operator is equivalent to simply reversing the bond signs along a semi-infinite path\cite{kadce}.
For the $\mu$-order operator the modified set $\{ J'_{ij} \}$ is still within the physical ensemble. 
Its average is therefore bounded below by zero but it can take any positive value\cite{readlud}, in contrast the order parameter 
which lies between $-1$ and $+1$.

Let us now consider the average of an observable $W(\{ J\} ,\beta)$. $W$ might represent
a two-point correlator as in Eq.(\ref{convcorr}) but is generally an arbitrary function over the bond configurations $\{ J\}$ and of
temperature. Ensemble averages will be denoted by an over-bar
\begin{equation}
\overline{W}=\int_{-\infty}^{\infty} W(\{ J\} ,\beta) \prod_{ij} P(J_{ij})dJ_{ij}.
\end{equation}
It was noticed by Nishimori\cite{nishiorg} that this average may be rewritten as
\begin{equation}\label{nishiavg}
\overline{W}= \text{Tr}_{\tau} \int_{0}^{\infty} W(\{ \tau J\} ,\beta)
Z( \{ \beta^{-1}\tau A\} , \beta) \prod_{ij}Q(J_{ij})dJ_{ij}
\end{equation}
where $Z(\{ \beta^{-1} \tau A\}, \beta)$ is an Ising partition function with a set of couplings
$\{ \beta^{-1}\tau_{ij}A_{ij} \}$ and
\begin{eqnarray}
Q(J_{ij})=[P(J_{ij})+P(-J_{ij})]/[2\cosh A_{ij}], \\
A(J_{ij})=-\frac{1}{2}\ln [P(J_{ij})/P(-J_{ij})], 
\end{eqnarray}
where $\tau_{ij}=\pm 1$. Thus the integral is taken over the positive
values of $J_{ij}$ only and the negative part of $P(J_{ij})$ is included through the sum over $\tau_{ij}$.
The Nishimori line is defined as the line on which $A(J_{ij})=\beta J_{ij}$, and hence 
$Z(\{ \beta^{-1} \tau A\}, \beta)=Z(\{ \tau J\}, \beta)$. 
Owing to this relation a number of averages can be performed exactly for bimodal and Gaussian distributions (and others)
on the Nishimori line.
In particular Nishimori\cite{nishiorg} calculated the internal energy in these two distributions exactly by considering the
average of the free energy, $\overline{\ln Z}$. The symmetry on the Nishimori line may also be expressed as a
replica symmetry\cite{doussal} from which further strong constraints on the distribution functions of correlators follow.

Now, for the $q$th moment of a disorder-dependent observable such as the correlation function in Eq.(\ref{altref}) 
on the Nishimori line one can write
\begin{equation}\label{gencorr}
\overline{\langle O_{x}O_{y}\rangle^q}=\text{Tr}_{\tau}\int_{0}^{\infty} \left( \frac{Z'}{Z}\right)^q Z \prod_{ij}Q(J_{ij})dJ_{ij}.
\end{equation}
From this it is evident that there appear symmetries in
the distribution of such correlation functions if there is a simple relation between $Z$ and $Z'$. 
In particular since any product of $\mu$-operators is
represented by reversing the sign of a set of bonds, Eq.(\ref{gencorr}) is invariant under the change $Z\to Z'$. 
This can be seen by noticing that taking the trace over $\tau$ 
inside the integral takes positive and negative bond strengths with equal weight. 
In turn, the change $Z\to Z'$ is equivalent to $q\to 1-q$ and hence the moments are symmetric with 
respect to $q=1/2$; The result $\overline{\langle O_{i}O_{j}\rangle^q}=1$ for $q=0,1$ readily follows from normalization
of Eq.(\ref{nishiavg}).

\section{Nishimori point of the $\pm J$-RBIM}\label{numerics}

The result of the previous section holds for the general RBIM with a Nishimori line. Here we restrict ourselves to
the $\pm J$ model with the bimodal bond distribution
$P(J_{ij})=p\delta(J_{ij}-1)+(1-p)\delta(J_{ij}+1)$ for which the Nishimori line is given by 
$\exp (-2\beta)=(1-p)/p$. On this line, the model has a disorder-dominated multicritical point known as the Nishimori
point. Its position has been estimated in the past and again, more accurately, recently\cite{merzch,honecker}. Two critical
exponents are now known from numerical calculations\cite{merzch}. We also focus on the Nishimori point here
and we consider the $q$th moments of the $\mu\mu$-correlator with separation $r$. We denote the averaged correlators
by $g(r;q)$. If they are critical with power law decay in the plane then
\begin{equation}\label{geom}
g(r;q)=\overline{\langle\mu_{0}\mu_{r}\rangle^{q}}=A (r/a)^{-\eta(q)},
\end{equation}
where $A$ and $a$ are constants and $\eta(q)$ is a $q$-dependent scaling
dimension. We would like to answer the following questions: Is this averaged correlator conformally invariant? 
If so, how are the $\langle\mu_{0}\mu_{r}\rangle$ distributed, or, equivalently, 
what is the functional dependence of $\eta(q)$ on $q$?  
As shown in the previous section $g(r;q)=1$ for all $r$ at $q=0$ and $q=1$. 
The scaling dimension $\eta(q)$ of $g$ has then two zeros in $q$. Furthermore $\eta(q)$ must be less than or equal to zero
for $q>1$ (and $q<0$) as shown by the following argument: we start from the inequality 
$\overline{x^p}\geq (\overline{x})^p$ for any real random $x$ satisfying $0\leq x \leq \infty$ and any real $p>1$. Applying 
this to $x=\langle\mu_{0}\mu_{r}\rangle$ one readily finds that
$\eta(p)\leq p\eta(1)$ and, since $\eta(1)=0$, the above result follows. Below we use numerical calculations to 
show that $\eta(q)$ is, in fact, non-zero and negative for $q>1$ and we determined its form in detail for a range of $q$.  

Conformal field theory predicts that the correlation functions in Eq.(\ref{geom}) decay on a cylinder with
circumference $M$ according to
\begin{equation}\label{conff}
\overline{\langle\mu_{0}\mu_{r}\rangle^{q}} = \left[ (M/a\pi)f(r/M)\right] ^{-\eta(q)},
\end{equation}
where
\begin{equation}
f(r/M)=\sin (\pi x/M)\ \ or\ \ \sinh (\pi y/M)
\end{equation}
and $r\equiv (x,y)$ with $x$ and $y$ the coordinate separations around the circumference and along the cylinder, respectively.
In addition one can also consider the logarithmic average (typical value) which is related to
the spectrum in $q$ through the derivative with respect to $q$ at $q=0$. If Eq.({\ref{conff}) holds then the typical value
decays on the cylinder as 
\begin{equation}\label{log}
\overline{\ln \langle\mu_{0}\mu_{r}\rangle}  = -\left(\left.\frac{\partial \eta(q)}{\partial q}\right| _{q=0} \right)
\ln [(M/a\pi) f(r/M) ].
\end{equation}

Using the mapping of the RBIM onto a network model of free fermions described in detail in Ref.\onlinecite{merzch},
and exploiting the efficiency of the transfer-matrix algorithm in the network formalism we calculate the correlation
functions in Eq.(\ref{conff}) and Eq.(\ref{log}).
The logarithmic average has small fluctuations compared to the direct averages 
and can be used as a sensitive test of whether the two-point correlators are actually conformally invariant.
Fig.(\ref{logplt}) shows the dependence of the typical value of the $\mu\mu$-correlator on cylinders with circumferences 
between $M=6$ and $M=22$ with separation along the cylinder $(x=0)$, calculated using 
$10^4$ samples. The statistical errors here are between $0.5$ and $1.0$ percent for
all system sizes and hence the error bars in the plot appear smaller than the symbols. The data fall accurately onto a 
single straight line with slope $k=-0.6911(17)$. 
The vertical dashed line is the point $r=M$ where, roughly, the correlator crosses over from two-dimensional to
quasi 1D behavior. The typical correlation function hence convincingly obeys the prediction from conformal invariance.

\begin{figure}[ht]
\begin{center}
\begin{tabular}{c}
\epsfig{file=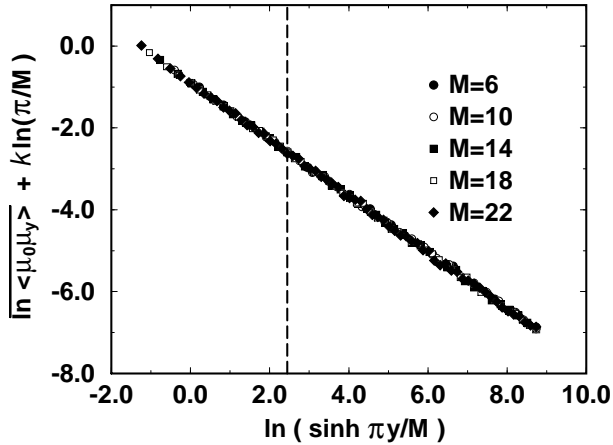,width=3.3in} 
\end{tabular}
\parbox{3.3in}{\caption{Scaling of the typical averages of the $\mu\mu$-correlation functions for $M=6-22$ where a 
straight line represents the form predicted by conformal invariance. The vertical
dashed line marks the crossover point from two-dimensional to quasi one-dimensional behavior. The statistical errors are between $0.5$ and $1.0$ percent throughout.
\label{logplt}}}
\end{center}
\end{figure}

The direct averages in Eq.(\ref{conff}) have extremely large fluctuations and are dominated by rare events. Therefore,
compared to computing the logarithmic average many more samples are needed. 
In the following we have typically used $10^8$ disorder realizations. We calculate the average
correlation functions on a cylinder now with the separation $r$ running around the circumference $(y=0)$ for $q$-values between $q=0$
and $q=2.5$; Four of the direct averages for $q=0.40,0.60,1.00,1.28$ and $M=22$ are displayed in Fig.(\ref{etasa}). Again, error bars are smaller than the symbol sizes but
now there is a strong system size dependence: the statistical errors range from $0.5$
percent for the smallest $q$ to $4.0$ percent for $q=1.28$.
The functional form dictated by Eq.(\ref{conff}) is, again, obeyed accurately and $\eta(q)$ can be read off as the
slope of the fitted lines. In Fig.(\ref{etasb}) the slopes for various values of $q$ are plotted versus $q$. As expected
the curve is symmetric with respect to $q=1/2$ and has a zero at $q=1$, crossing to negative values for $q>1$. The fact that
we reproduce $\eta(1)=0$ accurately provides a strong check of our numerical calculations, since this is a property that
emerges only after averaging over random bond realisations. 

\begin{figure}[ht]
\begin{center}
\begin{tabular}{c}
\epsfig{file=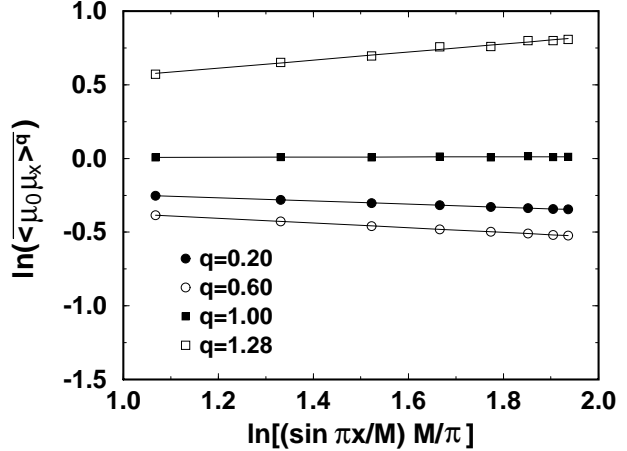,width=3.3in} 
\end{tabular}
\parbox{3.3in}{\caption{Scaling of the correlation functions for $M=22$ and four values of $q$. Here the separation $x$ is around the circumference and $y=0$. The statistical
errors range from $0.5$ percent for $q=0.20$ to $4.0$ percent for $q=1.28$
\label{etasa}}}
\end{center}
\end{figure}

\begin{figure}[ht]
\begin{center}
\begin{tabular}{c}
\epsfig{file=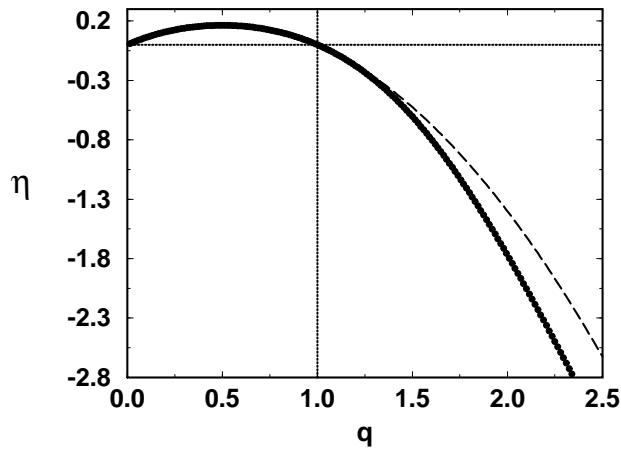,width=3.3in} 
\end{tabular}
\parbox{3.3in}{\caption{ The scaling dimension of the $q$-moment of the $\mu\mu$-correlator as a function
of q. The dashed line shows the approximate parabolic spectrum valid for $0.2<q<1.2$. For larger $q$ the
deviation is significant.\label{etasb}}}
\end{center}
\end{figure}

Our error margin for $\eta(q)$ shown in Fig.(\ref{etasb}) is less than three times the symbol sizes for $q>1$ but smaller than this
for $q<1$. Since the direct averages are dominated by rare events,  
under-sampling must be a concern. For higher $q$ this practical problem becomes worse. In order to assess
sample-to-sample fluctuations we have therefore repeated the data analysis while omitting different subsets of the 
data: through this we arrive at the above error margin.

Differentiating the curve $\eta(q)$ numerically at $q=0$ in order to compare the result with the typical average gives 
$\partial \eta/\partial q \approx 0.70(1)$ which agrees with the above value of $k$. 
On a cylinder of circumference $M$ this part of the multi-scaling spectrum may also be compared to previous results in 
the limit of infinite separation.
For the $\mu\mu$-correlator in the Ising model this limit amounts to the insertion of
an infinite seam of reversed bonds along the cylinder. Then, since the logarithmic average of $Z$ is simply the 
free energy, the typical value of the infinite $r$ $\mu\mu$-correlator $Z'/Z$ is the free energy difference
between periodic and anti-periodic boundary conditions. The free energy per site $\Delta f$ is the interfacial tension
which conformal field theory predicts to be $\alpha\pi M^{-2}$ with $\alpha$ a universal scaling amplitude. 
For the $\pm J$ RBIM at the Nishimori point it was shown by independent numerical calculations\cite{merzch,honecker,tom} 
that $\Delta f\approx 0.691(2) \pi M^{-2}$. This agrees within errors with the derivative at $q=0$ and coincides very well
with $k=-0.6911(17)$ obtained above.

In the context of multifractality it is interesting to know whether the spectrum in Fig.(\ref{etasb}) is piecewise exactly parabolic.
Such behavior is known to occur for the inverse participation ratios of the ground state wave-functions of fermions in a random
gauge potential\cite{ludfishank} and it has been proposed and confirmed numerically for wave-functions at the integer
quantum Hall transition\cite{ferd}. Parabolic spectra stem, in two dimensions, from quantities whose logarithms have the
distribution of a free massless field, and it can be shown that this would be 
compatible with the constraints on the Nishimori line. However, while
for $q<1.2$ the $q$-dependence of $\eta$ on $q$ seems indeed compatible with a parabola, at larger $q$ the curve 
$\eta(q)$ versus $q$ deviates significantly and an exactly parabolic spectrum at the Nishimori point can thus be excluded. This
is shown in Fig.(\ref{etasb}) where the dashed line is a parabola fitted to $\eta(q)$ between $q=0$ and $q=1$.

\section{Conclusions}

We have demonstrated the presence of negative 
scaling dimensions at the multi-critical point in the $\pm J$ random-bond Ising model by 
considering the the Kramers-Wannier dual operator. 
Since the latter it is not bounded from above in the presence of antiferromagnetic bonds 
its correlation functions can increase with separation in this case which we have demonstrated  
numerically for the $\pm J$ random-bond Ising model in two dimensions. 
Furthermore at the disorder dominated multi-critical point
in this model we find that the $q$th moments of the two-point correlator in the ensemble distribution
behave according to the prediction from conformal invariance and that, for $q>1$, the exponents $\eta(q)$ become negative. 
We have calculated the multiscaling spectrum $\eta(q)$ and we support 
our observations using symmetry properties of the Nishimori
line on which the critical point is located and which forces $\eta(0)=\eta(1)=0$.

\section{Acknowledgements}

We would like to thank N.~Read and J.~L.~Cardy for useful discussions.
The work in this paper was supported in part by EPSRC under grant GR/J78327.

\end{document}